# Towards order of magnitude X-ray dose reduction in breast cancer imaging using phase contrast and deep denoising

Ashkan Pakzad[1], Robert Turnbull[2], Simon J. Mutch[2], Thomas A. Leatham[3], Darren Lockie[4], Jane Fox[5,6], Beena Kumar[5,6], Daniel Häsermann[7], Christopher J. Hall[7], Anton Maksimenko[7], Benedicta D. Arhatari[7], Yakov I. Nesterets[8], Amir Entezam[1,7], Seyedamir T. Taba[3], Patrick C. Brennan[3], Timur E. Gureyev[1,9], Harry M. Quiney[1]

[1] *School of Physics, The University of Melbourne, Parkville, VIC 3010, Australia*
[2] *Melbourne Data Analytics Platform, The University of Melbourne, Parkville, VIC 3010, Australia*
[3] *Faculty of Health Sciences, The University of Sydney, Lidcombe, NSW 2141, Australia*
[4] *Maroondah BreastScreen, Eastern Health, Ringwood, VIC 3134, Australia*
[5] *Monash Health, Clayton, VIC 3168, Australia*
[6] *Faculty of Medicine, Nursing and Health Sciences, Monash University, Clayton, VIC 3800, Australia*
[7] *Australian Synchrotron, ANSTO, Clayton, VIC 3168, Australia*
[8] *Commonwealth Scientific and Industrial Research Organisation, Clayton, VIC 3168, Australia*
[9] *School of Physics and Astronomy, Monash University, Clayton, VIC 3800, Australia*

## Abstract

Breast cancer is the most frequently diagnosed human cancer in the United States at present. Early detection is crucial for its successful treatment. X-ray mammography and digital breast tomosynthesis are currently the main methods for breast cancer screening. However, both have known limitations in terms of their sensitivity and specificity to breast cancers, while also frequently causing patient discomfort due to the requirement for breast compression. Breast computed tomography is a promising alternative, however, to obtain high-quality images, the X-ray dose needs to be sufficiently high. As the breast is highly radiosensitive, dose reduction is particularly important. Phase-contrast computed tomography (PCT) has been shown to produce higher-quality images at lower doses and has no need for breast compression. It is demonstrated in the present study that, when imaging full fresh mastectomy samples with PCT, deep learning-based image denoising can further reduce the radiation dose by a factor of 16 or more, without any loss of image quality. The image quality has been assessed both in terms of objective metrics, such as spatial resolution and contrast-to-noise ratio, as well as in an observer study by experienced medical imaging specialists and radiologists. This work was carried out in preparation for live patient PCT breast cancer imaging, initially at specialized synchrotron facilities.

## Introduction
Breast cancer is one of the most prevalent cancers in the world and the most common in women globally (1, 2). Though breast screening programmes have been successful in identifying many sub-types of cancers at an early stage, challenges still remain (3).

X-ray digital mammography (DM) and digital breast tomosynthesis (DBT) are currently the mainstay of breast screening. In DM, 2D flat images are acquired. Details can be obscured, or artificial contrast can be created due to overlapping breast tissues, particularly in the case of dense breasts. Since cancers are of similar density as fibroglandular tissue and may appear similarly in imaging, they can be harder to detect in dense breasts (4), with the possible exception of microcalcification with morphology suggestive of intra-duct pathology or a malignant mass which have significantly higher density than soft tissues. In DBT, projection X-ray images are acquired within a limited angular range to produce a near-3D image, a step towards alleviating the overlapping tissue problem. Though DBT has shown improvement at detecting smaller lesions, its recall rate is still high in women with dense breasts (5). Furthermore, both DM and DBT involve forced breast compression to optimise the imaging conditions which can be uncomfortable and can discourage women from participating in screening programmes (6). Ultrasound and magnetic resonance imaging (MRI) serve as alternative modalities; however, both have a lower specificity than DM, often leading to further follow-up examinations (7).

X-ray computed tomography (CT) produces full 3D volumetric images, overcoming issues of tissue overlap, while also eliminating the need for compression. A breast CT image is acquired by capturing X-ray projections during a 360-degree rotation. These projections are reconstructed into a complete 3D image. The radiation dose in CT imaging can be reduced (at the cost of increased noise) by either lowering the dose per projection or decreasing the number of projections. However, insufficient numbers of projections limit the data available for accurate reconstruction, thereby compromising spatial resolution (8).

The breast is a highly radiosensitive organ (9) and radiation doses are typically reported as mean glandular dose (MGD). While increasing the dose reduces image noise, it raises the risk of radiation induced malignancies. The United States Food and Drug Administration (FDA) sets a limit of 3 mGy MGD per view for DM and DBT. Since standard screening requires two views, and a single CT would replace both, the theoretical upper limit for a single screening examination is therefore 6 mGy. Recent studies in DM and DBT report average doses in the range of 1.1 – 2.3 mGy and 1.6 – 2.5 mGy per view, respectively (10–13). Radiation doses are higher in breast CT screening, where commercially available systems deliver doses in the range of 5.1 – 7.2 mGy (14–16). It can be highly beneficial to leverage the combination of X-ray phase-contrast CT (PCT) and artificial intelligence (AI) techniques to reduce radiation doses as much as possible while maintaining overall image quality.

Conventional clinical X-ray imaging of the breast (DM, DBT and CT) rely on capturing differences in X-ray absorption by different tissues. PCT imaging additionally captures phase shifts due to interactions of X-rays with tissues. The additional phase information can be exploited to yield images with higher quality than in attenuation-based imaging at the same doses, or with equivalent image quality at lower radiation doses (17).

There are several methods to achieve X-ray phase-contrast imaging, but propagation-based phase-contrast imaging is the only technique that doesn't require special optical elements (18). In propagation-based imaging, the spatially coherent X-ray beam interacts with the object and then propagates over a relatively long distance before detection. The so-called homogenous transport of intensity equation algorithm (19) can be used to retrieve the changes in phase from the detected images. This algorithm significantly increases both the signal-to-noise ratio (SNR) and the contrast-to-noise ratio (CNR) compared to conventional absorption-based images collected at the same X-ray dose and spatial resolution (20, 21). The resultant image after phase retrieval and CT reconstruction can be expressed in terms of the imaginary part of the refractive index, $\beta$, which is related to the linear attenuation coefficient, $\mu$, by $\beta = \mu\lambda/(4\pi)$, where $\lambda$ is the wavelength of incident X-rays. In turn, $\mu$ is related to the local mass density via the mass-attenuation coefficient which is specific to the tissue and the X-ray energy. It is convenient to express the reconstructed images in $\beta$ as it is a dimensionless property that depends only on the tissue and the X-ray energy (wavelength). The approximate mean values for $\beta$ in glandular and adipose tissue at 32 keV X-rays used here are $1.1 \times 10^{-10}$ and $8.1 \times 10^{-11}$, respectively.

As PCT yields a higher CNR in breast imaging (22), phase-contrast images can be acquired at a lower dose to yield the same image quality, already qualifying it as superior to attenuation-contrast imaging. The widespread introduction of propagation based imaging in medical breast imaging is primarily impeded by the requirement of high spatial coherence of the X-rays suitable for this technique, which is currently only achievable with synchrotron or microfocus sources, with the former being quite expensive and the latter currently not being bright enough for clinical imaging (23).

Reducing radiation dose to ameliorate the risks of X-ray imaging of the breast can result in increased image noise. This noise can be reduced through post-processing, known as denoising. For example, one could bin multiple detector pixels which would effectively increase the number of registered photons at each pixel, increasing the SNR. This would come, however, at the expense of spatial resolution. The latter leads to a decrease of image sharpness and a reduction in the visibility of small features, such as breast microcalcifications. This trade-off between noise and spatial resolution, $res$, for the same sample and dose gives rise to the known noise-resolution duality (24). In the case of 3D imaging, this essentially states that the ratio $SNR^2/res^3$, is proportional to the dose. This fact can be understood by recognising that both the dose and $SNR^2/res^3$ are proportional to the mean number of photons that interacted with the unit volume of the imaged sample during the scan. While SNR can be raised using low-pass filters, $SNR^2/res^3$ is invariant with respect to linear image filtering operations (17). It is difficult, therefore, to increase SNR without spoiling the spatial resolution or increasing the dose, highlighting that denoising is not a trivial task.

The importance of image clarity has resulted in numerous methods and approaches to denoising medical images, with deep learning models consistently demonstrating superior performance (25). In supervised deep denoising, noisy images are fed to a deep learning model which learns a mapping to a reference noiseless (or less noisy) image. Importantly, deep learning is a non-linear process which implicitly utilises *a priori* knowledge and, therefore, it is not bounded by the same noise-resolution duality as linear filtering methods.

Previous studies in deep denoising of breast CT images have used cone-beam CT, training AI models to recover images from sparse subsets of projections, using reconstructions from the full set of projections as ground truth (26–28). Our study extends beyond prior work in the following respects: (i) we use PCT, which enables imaging at lower doses while achieving better image quality than in absorption-based CT; (ii) we use fresh mastectomy specimens, allowing for higher radiation doses than possible with live patients, to obtain high-quality ground truth images; (iii) we evaluate denoising on images reconstructed from all available projections, thereby avoiding the deterioration of spatial resolution associated with limited-projection data (8); (iv) we demonstrate improved reference-less (i.e. SNR) objective image quality after AI denoising; and (v) we confirm that these improvements are meaningful for diagnostic performance. We demonstrate that the combination of PCT with AI denoising can provide an order of magnitude or more dose reduction in PCT breast imaging without any loss of objective and subjective (according to visual assessment) image quality. Our PCT scans were collected at the clinically acceptable dose of 4 mGy MGD and a voxel size of 99 $\mu m^3$ in the reconstructed images. This study was carried out in preparation for live patient breast cancer imaging at the Australian Synchrotron.

## Results

In total, 34 mastectomy samples were imaged, (median ± interquartile range) donor age was (50.5 ± 18.75) years, weight (367.5 ± 276) g, and one was from a male donor. Mastectomy samples ranged in presentation. To summarise histopathology reporting, 18 were benign and 16 showed malignancies with tumours graded in range I to III. Of the malignant cases, 10 were invasive carcinomas of no special type (NST), 8 were ductal carcinoma *in situ* (DCIS) and 2 were invasive lobular carcinomas (ILC). One case had visible seams in axial slices where projections were stitched; this was deemed unsuitable for visual assessment and therefore excluded. Two PCT scans of each sample were collected, at 4 mGy and 24 mGy MGD, using monochromatic parallel 32 keV X-ray beam.

### Objective image quality metrics

Visual examples of denoised images are shown in Figure 1. Median SNR of 4 mGy, 24 mGy and denoised images were 7.4, 17.0 and 29.4 in adipose tissue, and 9.1, 21.1 and 34.5 in glandular tissue, respectively. Median contrast (see Methods) was 0.17 in 4 mGy images and 0.15 in both 24 mGy and denoised images. However, accounting for noise, median CNR measured at 1.3, 2.8 and 4.6 for of 4 mGy, 24 mGy and denoised images respectively. These results are shown in Figure 2.

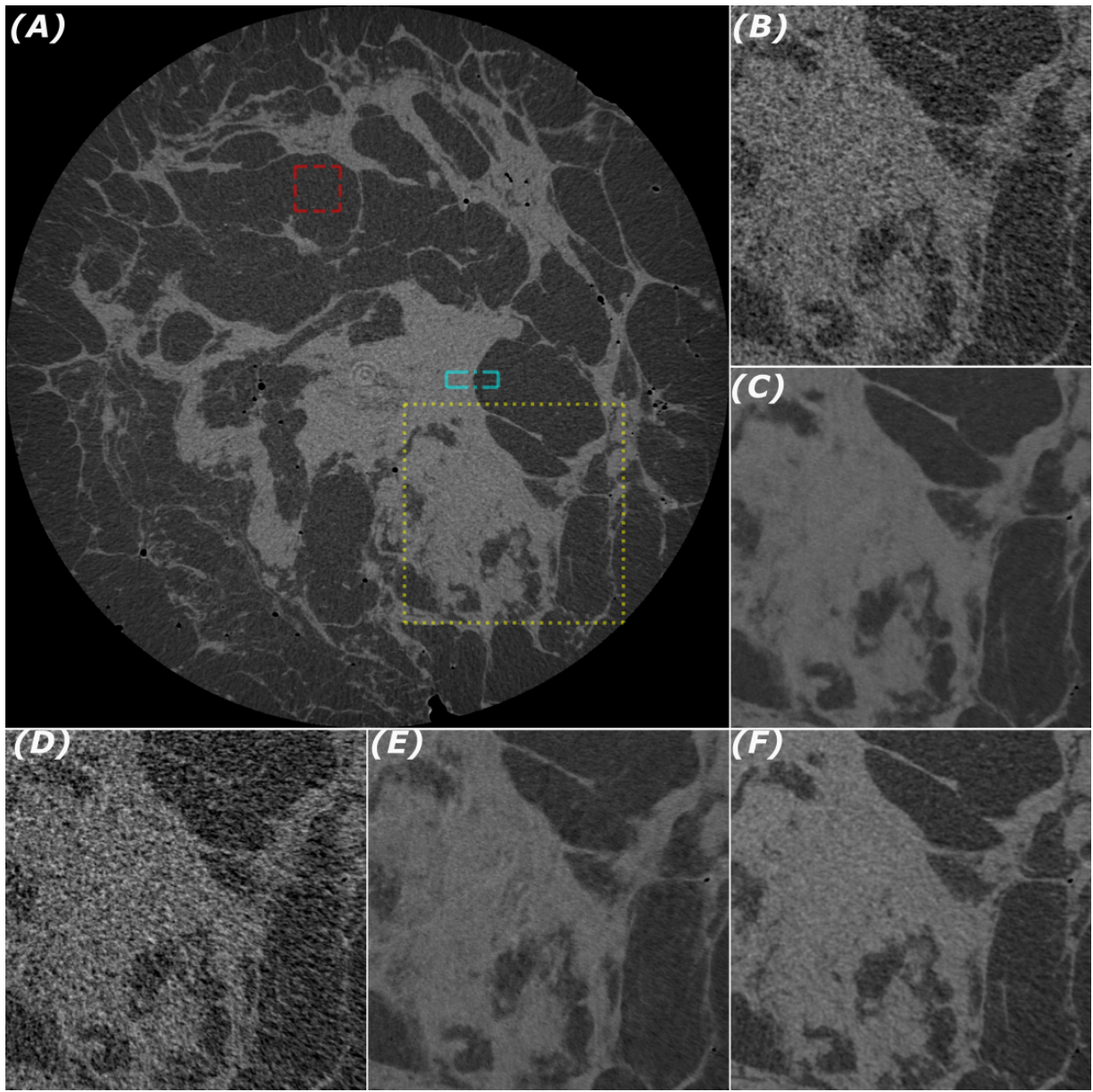

*Figure 1 Coronal slice of X-ray phase-contrast computed tomography (PCT) imaged breast from test-set. A, full view of image acquired at 24 mGy. Red dashed patch is an example of a homogenous adipose tissue patch for computing reference-less signal-to-noise ratio and resolution. Cyan dot-dash patch indicates example of glandular-adipose tissue boundary patch for measuring contrast and contrast-to-noise ratio. Yellow dot patch indicates magnified crop area of B-F. B is from the original 4 mGy scan, and C is after denoising. D is the half-subsampled reconstruction of B, i.e. 2 mGy image, and E is after denoising. F is the 24 mGy reference, however, note that B-E and F are from two different PCT scans of the same sample.*

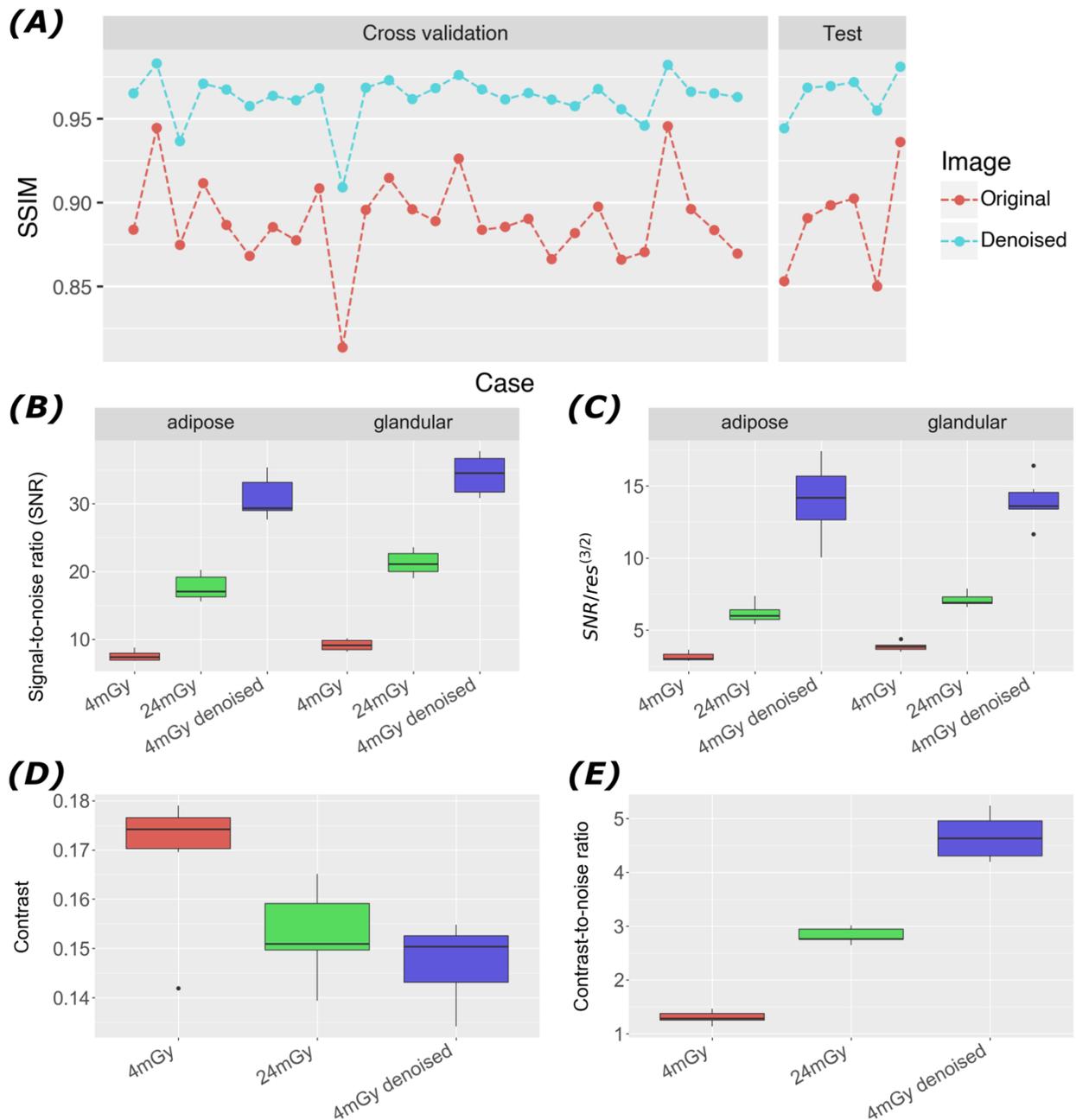

*Figure 2 A, Full-reference structural similarity index measure (SSIM) of original 4 mGy scan and its denoised result with respect to the reference 24 mGy scan for all cases. B-E are reference-less Image quality measurements made on homogenous and interfacing patches of adipose and glandular tissue (as explained in the main text) in test cases. The label $res$ denotes resolution.*

Median spatial resolution (see Methods) was found to be the same in the original (180 $\mu m$) and denoised 4 mGy images (180 $\mu m$), but was slightly worse in 24 mGy images (200 $\mu m$). The $SNR/res^{3/2}$ ratio was 4 times higher in denoised images than in the original 4 mGy ones (Figure 2), which was due to the corresponding increase of SNR after denoising, while the spatial resolution was unchanged. Note that the ratio of measured values of $SNR^2/res^3$ in 24 mGy and the original 4 mGy images was in good agreement with the ratio of their doses, i.e. 24 / 4 = 6, as predicted by the theory.

Registered 4 mGy images were evaluated with respect to 24 mGy images in terms of structural similarity index measure (SSIM) (29) as shown in Figure 2. Mean SSIM per image was 0.89 before denoising and improved to 0.96, so that the AI denoising increased the "similarity" between the 4 mGy images and the 24 mGy images of the same samples.

## Visual assessment

The results of the subjective assessment are summarised in Table 1. Based on Intraclass correlation coefficient (ICC) (30) assessors displayed excellent reliability for image noise (ICC=0.94) and overall image quality (ICC=0.81), good reliability for artifacts (ICC=0.63) and fair reliability for calcification visibility (ICC=0.52). However, there were contradictions in ratings between assessors for perceptible contrast (ICC=−0.41) and sharpness of tissue interfaces (ICC=−0.20).

Evaluation of the area under the visual grading curve $AUC_{VGC}$ (31) found that, after denoising, the 4 mGy images were rated significantly better in terms of image noise ($AUC_{VGC}$=0.91) and overall image quality ($AUC_{VGC}$=0.77) compared to the original 4 mGy images. Compared to the 24 mGy images, the denoised images were rated significantly better for image noise ($AUC_{VGC}$=0.81). With respect to the presence of artifacts, when rated against the 24 mGy reference, the 4 mGy images were rated lower both before ($AUC_{VGC}$=0.31) and after ($AUC_{VGC}$=0.31) denoising. For all other criteria, the denoised images were rated as equivalent to the 24 mGy reference, with no statistically significant differences.

*Table 1 Summary of visual assessment results. Asterix indicates statistical significance (\*=p<0.0167, \*\*=p<0.001) ICC = intraclass correlation coefficient; CI = confidence intervals; $AUC_{VGC}$ = Area under the visual grading curve*

|  |  | Average image quality score | | $AUC_{VGC}$ (95% CI) | | |
|---|---|---|---|---|---|---|
| Image quality criteria | ICC | 4 mGy | 4 mGy Denoised | 4 mGy against 24 mGy | 4 mGy denoised against 24 mGy | 4 mGy denoised against 4 mGy |
| Perceptible contrast | -0.41 | 0.04 | -0.11 | 0.52 (0.36,0.69) | 0.44 (0.35,0.51) | 0.48 (0.29,0.67) |
| Sharpness of tissue interfaces | -0.20 | -0.01 | -0.20 | 0.49 (0.32,0.66) | 0.41 (0.27,0.54) | 0.48 (0.25,0.65) |
| Image noise | 0.94 | -0.99 | 0.63 | 0.09 (0.01,0.20)** | 0.81 (0.69,0.91)** | 0.91 (0.80,0.99)** |
| Calcification visibility | 0.52 | 0.10 | 0.12 | 0.55 (0.49,0.60) | 0.55 (0.50,0.61) | 0.51 (0.46,0.56) |
| Artifacts | 0.63 | -0.50 | -0.46 | 0.31 (0.20,0.42)* | 0.31 (0.19,0.42)* | 0.51 (0.47,0.58) |
| Overall image quality | 0.81 | -0.79 | -0.07 | 0.15 (0.06,0.25)** | 0.46 (0.33,0.59) | 0.77 (0.61,0.90)** |

As PCT images are based on β values, a physical quantity, the contrast between different tissue areas does not change with dose. However, a negative ICC which indicates systematic disagreement between visual observers was found for two criteria, perceptible contrast and sharpness of tissue interfaces. One explanation for this is that the presence of noise can introduce a variation, an impression of fine-detail contrast in an image. This may be explained by the objective measures of contrast and SNR where an anti-correlation was found. The appearance of greater noise across a tissue boundary results in more extreme values in the image, therefore, increasing the gap in the average maximum and minimum values which corresponds to a greater perceived contrast.

Visual assessment results indicate calcification visibility was preserved in denoising. From a clinical perspective, the presence of calcifications may indicate malignant growth. Since they appear extraordinarily bright in X-ray images due to being highly attenuating, their detection is relatively easy. However, the shape of a microcalcification is important, and can only be assessed if the spatial resolution is sufficiently high. It is therefore important that the spatial resolution was preserved in our AI denoising.

### Lower dose denoising

The present study was focussed on AI denoising of 4 mGy PCT scans of mastectomy samples. However, following the successful denoising of 4 mGy dose images, we also evaluated the denoising model on 2 mGy test images. The original 4 mGy scan was subsampled by using half (2400) of the projections to reconstruct the effectively 2 mGy scan, the visual result of which is included in Figure 1. Similar improvements to the 4mGy case were observed here, with 16 times gain in $SNR^2/res^3$ after denoising. This preliminary evaluation of deep denoising of subsampled 2 mGy scans was conducted in view of a possible future extension of the present work aimed at finding the lowest possible dose at which the denoised breast PCT images would still be considered suitable for clinical radiological purposes.

## Discussion and conclusions

To our knowledge, this is the first demonstration of deep denoising of phase-contrast CT images of human breast tissue. The primary comparison between 4 mGy images with and without denoising demonstrated substantial improvements following denoising in terms of visual and objective metrics.

Phase-contrast X-ray imaging has already demonstrated an improved SNR and CNR compared to conventional attenuation contrast (22). In this paper, we have further amplified this technology by incorporating deep denoising models. Our results showed an improvement in median SNR by a factor of 4, without loss of spatial resolution, by denoising 4 mGy images. Since SNR is proportional to the square root of dose for the same samples, i.e. $SNR \propto \sqrt{Dose}$, this corresponds to an effective 16-fold dose reduction without a loss of image quality. It is critically important that this increase in SNR and CNR was achieved without any loss of contrast or spatial resolution. This is directly reflected in the increase of the CNR and $SNR/res^{3/2}$ metric after denoising. The ratio $SNR^2/res^3$, in particular, describes the amount of Shannon information available in the image (24). The fact that this information has apparently increased after denoising, seemingly violates the noise-resolution uncertainty relation, as it stands. Indeed, the $SNR^2/res^3$ ratio increased 16-fold after denoising without an increase in dose. That is, it appears that more information was obtained from the same number of photons than is physically possible. This can be explained by introducing an additional *a priori* information term, as the deep denoising model acts in the capacity of a mapper between low and high dose images.

We emphasise that we used a discriminative AI model that learns direct relationships between low and high dose images, unlike a generative model that is indirect and learns data distributions. Learning direct mappings reduces the likelihood that discriminative models will introduce artifacts. Images before and after denoising were rated equivalent by visual assessors with respect to artifacts, indicating that features present in 4mGy images were preserved and no new features were spuriously inserted by the AI model.

Our final trained models are deployable to standard consumer-grade machines, as GPUs are not necessary at the model application (inference) stage, and denoised images are available within a few minutes. Though the model was not trained on lower doses, the evaluation in denoising 2 mGy images continued the same trend without spoiling spatial resolution. This indicates that even lower doses than currently used in DM and DBT can be investigated in conjunction with deep denoising in the future.

Previous breast CT denoising studies used an FDA-approved breast CT scanner (26–28), with a voxel size of 273 μm$^3$, compared to 99 μm$^3$ in this study. Since the radiation dose in conventional CT is inversely proportional to the fourth power of spatial resolution (32), this difference corresponds to an approximately 58-fold reduction in dose in our case $(273/99)^4 \approx 58$. High spatial resolution is also crucial for adequate visualisation of microcalcifications.

These results are particularly important for breast cancer imaging of high-density breasts. It should be noted that the problem for individuals with radiologically dense breasts are three-fold. (i) The low contrast between breast tissues, means greater obscurity with the presence of more glandular tissue. An increase in dose would improve this. (ii) Glandular tissue is radiosensitive and, therefore, its exposure to ionising radiation should be limited as much as possible. (iii) Epidemiological studies have shown that women with dense breast tissue have a higher risk of developing breast cancer (33). The results presented here are of potential clinical significance, since the demonstrated image quality improvement via a combination of phase-contrast imaging and deep denoising could potentially yield greater diagnostic specificity with the radiation dose reduced by an order of magnitude.

The most significant limitation of X-ray phase contrast in breast imaging is that it has so far only been demonstrated using synchrotron sources (20, 22, 34). Using smaller microfocus X-ray sources for lab-based imaging would significantly increase the potential for clinical use and is therefore an active area of research. The main limitation of microfocus X-ray sources for medical imaging at present is their low brightness, which is insufficient to complete a breast scan within a reasonable time such as e.g. 10-15 seconds that the patient could be expected to hold their breath. As the AI denoising allows one to obtain images with radiologically relevant quality at much lower doses, it opens the possibility to complete a CT scan within a much shorter time, thus considerably lowering the threshold for the introduction of microfocus X-ray sources into medical PCT imaging.

Both generative (23, 35, 36) and unsupervised (37) models have recently demonstrated success in CT denoising literature and subsequent research should consider comparisons with these approaches. By scanning mastectomies, we can use high doses, such as 24 mGy MGD, which of-course is not possible in the case of live patients. This technology has the potential to generalise well to in-patient scans, the use of fresh mastectomy samples means that these images should resemble data obtained from live patients. Anatomical anomalies related to the nature of mastectomy samples such as voids are small and infrequent enough to be masked out from the process of model training. Therefore, this technology could also be used in the future to optimise images in conventional breast CT, DBT and mammography by utilising a similar approach.

## Methods

Ethical approval was received from the Human Research Ethics Committee (project number: CF15/3138 2015001340). All participants that donated mastectomy samples for the purpose of research provided consent prior to surgery. Participants underwent a mastectomy operation for the clinical purposes of removing underlying malignancies or reducing the risk of developing malignancies.

### Samples and scan acquisitions

Mastectomies were donated and scanned by PCT within hours of excision and without any fixation or forced compression. After scanning, samples were sent for histopathological analysis and reporting. For the scans, mastectomy samples were placed in nipple up orientation in a thin-walled cylindrical container of 11 cm diameter. Samples were consistently orientated using surgical sutures and clips.

All scans were performed at the Imaging and Medical Beamline (IMBL) facility of the Australian Synchrotron. A parallel monochromatic X-ray beam was generated at the energy $E = 32$ keV and energy resolution $\Delta E / E \cong 10^{-3}$ illuminating the detector with a cross-sectional area of 150 (horizontal) x 37 (vertical) mm$^2$. Separate scans of each mastectomy at 4 mGy and 24 mGy MGD were acquired (by changing photon flux only) at a source-to-sample distance of 137 m and a 6 m free-space propagation distance from sample to detector. Projection images were captured on a Teledyne-Dalsa Xineos-3030HR flat panel detector with an active area of 296 × 296 mm$^2$ and pixel pitch of 99 μm. Each scan consisted of 4800 projections over 180 degrees with an angular step of 0.0375 degrees.

An ionisation chamber was used to measure the delivered dose to air. Associated MGD was subsequently calculated using Monte Carlo simulations described in (38). Samples that were larger compared to the beam height had vertically displaced CT scans captured with overlap that were later stitched. Flat field and dark images were also acquired without any sample and in the latter case without radiation in order to correct sample projections. Three-dimensional distributions of $\beta$ in the mastectomy samples were reconstructed from the CT scans using the Unified Tomographic Reconstruction (UTR) method (39) with Paganin's phase retrieval algorithm (19). The original 4 mGy scan was then rigidly registered with the 24 mGy scan by minimising mean square difference using ITKElastix (Version 0.21.0) (40).

### Denoising model

We trained a U-Net to denoise the acquired 4 mGy PCT breast mastectomy scans. This was achieved by training on equivalent 4 mGy scans produced by subsampling the 24 mGy scans. The "subsampling" means that these equivalent 4 mGy training scans were produced by using one sixth (800 projections) equally spaced projections from the 24 mGy scan (with 4800 projections). The set of projections in each subsampled training reconstruction was mutually exclusive, resulting in six different representations of noise per each 24 mGy scan. The trained model could then be evaluated on the original 4 mGy scans reconstructed from a full set (4800) of projections.

Six cases were held out from the training process for the purposes of final testing, the remaining 28 were split into five partitions in a five-fold cross-validation scheme. Therefore, five separate denoising models were trained, each validated on exclusive sets of images. Final test cases were denoised using the model that yielded the lowest validation loss.

Figure 3 shows a schematic of model training and evaluation in this study.

We used a 3D U-Net architecture (41) due to its proven success and wide adoption for medical image tasks. The objective function was set to minimise the smooth L1 loss (which uses L2 when loss falls below 1). The model was trained to predict the noise residual image of low dose input, i.e. difference with the 24 mGy image. Original calcification voxels ($\beta \geq 2 \times 10^{-10}$) were used to replace denoised voxels in the final output.

Model training was facilitated through use of TorchApp (Version 0.3.10) (42) with a fastai (Version 2.7.14) backend (43), an abstraction of PyTorch. AI model training was done using an Intel(R) Xeon(R) Gold 6326 CPU @ 2.90GHz processor with Nvidia A100 GPU card, taking an average of 33 hours training time per cross-validation partition. Source code is publicly available at https://github.com/quell-devs/quell.

### Objective evaluation metrics

We evaluated image pairs using SSIM, which can vary between 0 and 1, where the latter indicates perfect similarity. In the computation of SSIM, the range of values $[5.0 \times 10^{-11}, 7.0 \times 10^{-10}]$ was linearly rescaled and clipped to $[0, 1]$.

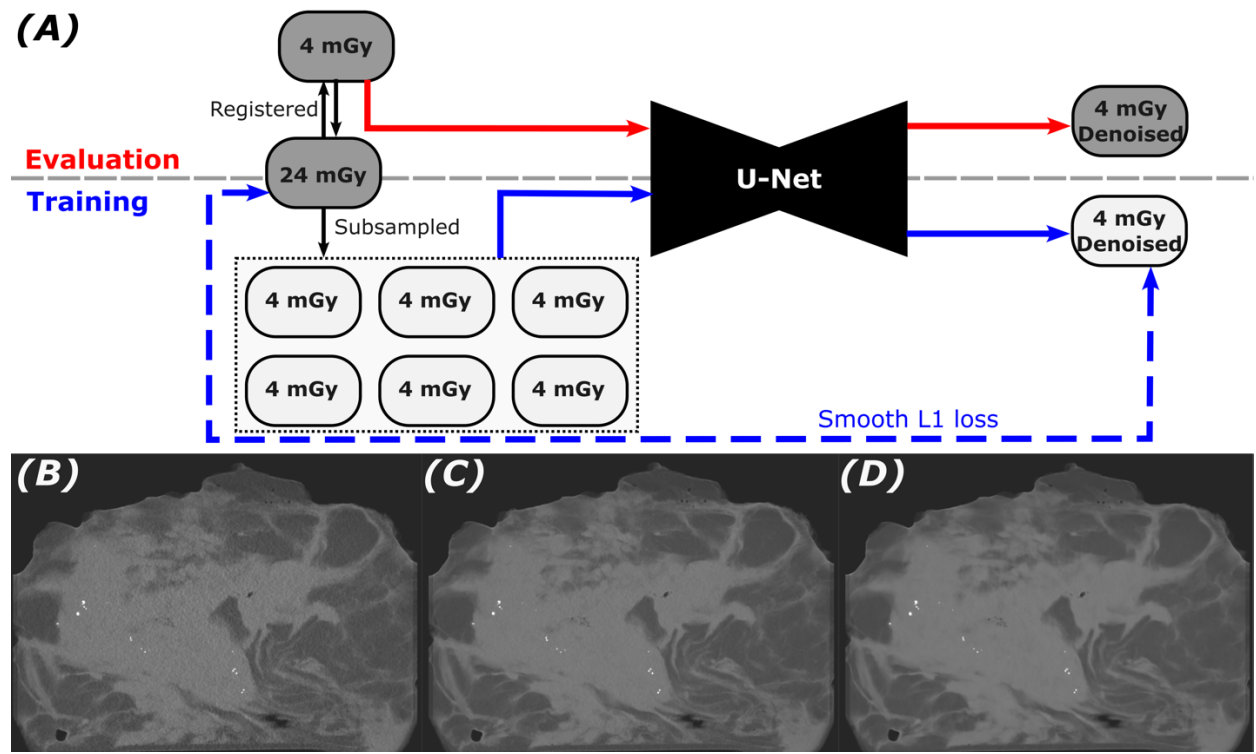

*Figure 3 A, Schematic demonstrating U-Net model training by subsampling projections of 24 mGy mean glandular dose (MGD) scans to produce six noise independent 4 mGy reconstructions. Note that evaluation of the model was based on denoising registered reconstructions of 4 mGy scans without subsampling (real world scenario). B, C and D are an example set of axial thick slice PCT images shown in the visual assessor study comparing 4 mGy, 24 mGy and 4 mGy denoised respectively. This case was in the test dataset, found to have a 135 mm, grade 3 ductal carcinoma in situ.*

SNR and spatial resolution were computed by analysing voxels in homogenous patches of the images. SNR was defined as $SNR = \bar{x}/\sigma$, where $\bar{x}$ is the average and $\sigma$ the standard deviation of the pixel values within the selected patch. Spatial resolution (*res*) was defined as the width of the point-spread function (PSF) of the imaging system. Assuming that the PSF is approximately gaussian, one can derive the resolution of the system using methods described in (44) by evaluating the width of the noise frequency spectrum in a flat area of an image. The two metrics were evaluated separately with respect to adipose and glandular tissues. Contrast was measured at glandular-adipose tissue interfaces by binning voxels in a rectangular selection across the interface based on $\beta$ value into five bins. Denoting the average top and bottom bin values as $\beta_{max}$ and $\beta_{min}$ respectively, $Contrast = (\beta_{max} - \beta_{min})/(\beta_{max} + \beta_{min})$. CNR was defined by, $CNR = Contrast/\sigma_{background}$, where $\sigma_{background}$ was the standard deviation of voxels in an adjacent homogenous adipose tissue patch to the measured interface. Reference-less metrics such as SNR, spatial resolution, contrast and CNR were evaluated using XLI software (45). Three patches from different coronal slices were evaluated and averaged for each metric per test image. Image patch examples are shown in Figure 1.

### Expert visual assessment

Eleven assessors systematically evaluated the 4 mGy dose images with and without denoising to the reference 24 mGy dose images. Four assessors in this study were breast specialist radiologists, while the remaining seven assessors were medical imaging scientists and diagnostic radiographers.

Thick axial slices with thickness of 3.0 mm and 1.5 mm overlap were prepared. A calcification mask was made by evaluating the threshold of $\beta \geq 2 \times 10^{-10}$, while excluding isolated voxels. Slice columns were averaged, where calcification voxels were present. All other voxels were ignored, enhancing the appearance of calcifications. Examples are shown in

Figure 3. Thick slicing acts as a linear denoiser at the cost of spatial resolution, therefore, the ratio $SNR^2/res^3$ remains unchanged in this reslicing.

Assessors were blinded to which images were denoised and rated each test-reference pairs by six different criteria. Assessments were conducted in conditions resembling digital mammography reading rooms, with the assessments performed on full-resolution diagnostic monitors using the DetectedX image viewing platform (46). The three images of the same case could be simultaneously viewed with synchronised scroll, pan and magnification tools.

Assessors were asked to evaluate the image quality for the test images in the left and right panels against the reference images using a five-point rating scale from -2 to 2. Assessors rated each test-reference pair using the following six image quality criteria. (i) **Perceptible Contrast**, the differences in radiolucency between soft tissues, particularly within higher-density tissue (which appear brighter than adipose tissue), indicating how well soft tissue variations were depicted. (ii) **Sharpness of Tissue Interfaces**, the clarity of boundaries between different tissue types, measuring how well the image visualised transitions between tissues. (iii) **Image Noise**, the degree of quantum mottle in the image, with a higher score indicating less noise in the test image compared to the reference. (iv) **Calcification Visibility**, the sharpness and visibility of micro-calcifications, specifically assessing their clarity and prominence in the image. (v) **Artifacts**, the presence of artifacts in the image, focussing on newly introduced features in the test images compared to the reference by assuming the reference image is the gold standard and free of those artifacts. Note that ring artifacts, which typically present as bright vertical lines in the axial view, were ignored in this assessment. (vi) **Overall Image Quality**, a holistic assessment of the entire CT image, considering all aspects of image quality.

Inter-observer agreement was measured by ICC, where a score closer to the maximum value of one, indicates greater agreement between observers and therefore greater reliability in the measure. An ICC less than 0.4 suggests poor agreement and a negative ICC indicates systematic disagreement between assessors. ICC was computed based on absolute rating scores using IBM SPSS statistics v29.

For each image quality criterion, the cumulative distributions of the rating scores for the test images were plotted against the reference images, yielding a visual grading curve (VGC). An $AUC_{VGC}$ of 0.5 indicates equivalence between the test and reference images, whereas (0.5,1.0] indicates the test images are of higher quality and [0.0,0.5] indicates the reference images are of higher quality. VGC analysis was carried out on the assessors scores using VGC Analyzer software v1.0.2 (31).

Bootstrapping by 2000 re-samplings of assessor scores were conducted to determine 95% confidence intervals and p-values were computed based on the bootstrapped differences. Since 3 comparisons were made per criterion, the significance level was adjusted to control for multiple tests, $\alpha = 0.05/3 = 0.017$.

Additional details regarding methods in this study can be found in the supplementary materials.

## Acknowledgements

We would like to thank all participants who donated their tissue, without whom this would not have been possible. This research was funded by the National Health and Medical Research Council [2021/GNT2011204].

# Supplementary

## Data preprocessing

We implemented a 'half phase' retrieval, setting the gamma parameter to 275, i.e. half of the theoretical ratio of the differences in imaginary and real decrement components of the complex refractive index between adipose and glandular breast tissue, for 32 keV X-rays. This was found to produce visually preferable images compared to 'full phase' retrieval (1). Both tuneable parameters of Unified Tomographic Reconstruction, Tikhonov regularisation and noise-to-signal factor were set to small values to have minimal influence on reconstruction, $1 \times 10^{-6}$ and of 0.01 respectively.

Reconstructed volumes were pre-processed to remove the image of the container using an 11cm diameter cylinder which corresponded to the diameter of the breast container. All reconstructed PCT data (originally reconstructed in terms of the true values of the imaginary part of the refractive index, β) were linearly rescaled to a standardised range such that $\beta = 5.0 \times 10^{-11}$ and $\beta = 7.0 \times 10^{-10}$ became 1 and 10, respectively. The rescaled data was converted to 16-bit floating-point precision to optimize memory usage and computational efficiency.

## Deep learning model architecture and training

We used a 3D UNet architecture (2) with three down blocks and three up blocks, with skip connections between them. Convolutional layers had a kernel size of 3x3x3 voxels. The model had a total of 5,808,353 trainable parameters. The objective function was set to minimise the smooth L1 loss (which uses L2 when loss falls below 1) between predicted and target residual images. The batch size was set to 2 and models were trained over 40 epochs. ADAM optimiser with 1cycle policy (3) was used with maximum learning rate 0.001. The lowest validation scoring epoch was used for testing. Training runs were monitored using Weights & Biases (Version 0.16.3) (4).

Denoising inference was conducted by using a sliding 3D window of the fixed input size (256x256x64) with an overlap of at least 32 voxels in every dimension. Prediction windows were weighted by distance to window edge and stitched by weighted interpolation. Original voxels corresponding to calcifications ($\beta \geq 2 \times 10^{-10}$) replaced denoised voxels in the final output.